\documentclass[12pt]{iopart}
\usepackage{amsfonts}
\begin{document}
\jl{6}
\title[ Polydimensional branes]{Fuzzy dimensions and Planck's 
Uncertainty Principle for $p$-branes}

\author{Antonio Aurilia\dag{}\footnote[1]{e-mail
        address: \texttt{aaurilia@csupomona.edu}},
Stefano Ansoldi\ddag{}\footnote[3]{e-mail
        address: \texttt{ansoldi@trieste.infn.it}},
        Euro Spallucci\ddag{}\footnote[5]{e-mail
        address: \texttt{spallucci@trieste.infn.it}}}
\address{\ddag{}Department of Physics\\
            California State Polytechnic University\break
            Pomona, CA 91768 }

\begin{abstract}
The explicit form of the quantum propagator of a bosonic $p$-brane,
previously obtained by the authors in the quenched-minisuperspace
approximation, suggests the possibility of a  novel, unified, description
of $p$-branes with  different dimensionality. The background metric 
that emerges in this framework  is a quadratic form on a Clifford 
manifold. Substitution of the Lorentzian metric with the Clifford line element
has two far reaching consequences. On the one hand, it changes the very
structure of the spacetime fabric since the new metric is built out of
a  \textit{minimum length} below
which it is impossible to resolve the distance between two points; on the
other hand, the introduction of the Clifford line element extends the usual
relativity of motion to the case of \textit{relative dimensionalism} of all
$p$-branes that make up the spacetime manifold near the Planck scale.
\end{abstract}

\pacno{11.17}
 
\submitted 
  
\maketitle

\section{Introduction}
A surface that appears smooth to the naked eye, looks rough when seen under
a microscope. With increasing power of resolution, that seemingly
$2$-dimensional surface may reveal its thickness and eventually its atomic
``graininess'', at which point the very definition of dimensionality
becomes fuzzy.\\
It has long been conjectured that the same concept applies to the spacetime
manifold: the short distance spacetime geometry and topology are subject to
violent quantum fluctuations \cite{wheeler}. By ``short distance'', in
this context, one usually means the Planck scale, and, by ``violent'', we
mean that the quantum fluctuations in the geometry are of the same order of
magnitude as the background metric, so that the smooth spacetime manifold
of classical physics breaks down into a quantum foam of virtual
gravitational bubbles \cite{hawking} and other ``extendons'', or
$p$-branes \cite{firm}. In this perspective, it is to be expected that 
quantizing the classical theory of gravitation, namely, General 
Relativity, is a hopeless task. At the Planck scale, where quantum 
effects are deemed to become relevant, spacetime may not even be  a 
primitive concept. Rather than being a pre--existing, albeit dynamical, 
background as in General
Relativity, spacetime is an ``emergent property'',  that is, a low energy,
effective construct of the truly fundamental objects it is made of. Let us
call ``$p$-cells'', as opposed to $p$-branes, the elementary simplexes
that constitute the \textit{  pre-geometric  quanta} of the brane world 
manifold. $p$-cells are model manifolds that form the basis of algebraic 
topology, and can be defined independently of any spacetime background. 
Physical spacetime emerges from a \textit{  mapping}, $X$, of $p$-cells 
into a target manifold. The \textit{image} of  a collection of $p$-cell 
under the mapping $X$ corresponds to a physical $p$-brane\footnote{Here and 
in the following,  by convention, ``$p$'' stands for the  order of the 
simplex, or the
\textit{  spatial} dimension of its image, so that a $p$-brane is really a
$(p+1)$-dimensional submanifold in spacetime.}. Thus, a  $0$-cell, that is
a point with no spatial dimensions, corresponds to a world--line in
spacetime; a $1$-cell, or line element, corresponds to a ``spacetime
strip'', or world--sheet element, and so on, all the way up to the maximal
rank $(D-1)$-brane enclosing an elementary spacetime $D$-volume.\\
By analogy with a superconducting medium, which consists of a condensate of
Cooper pairs, classical spacetime may be regarded as a sort of string or 
brane condensate \cite{ume}, \cite{ume2}. Near some critical scale, 
presumably the Planck energy, the spacetime manifold
``evaporates'' into its fundamental constituents, i.e., strings and branes of
various dimensionality. In this extreme phase, spacetime consists of a
``gas of branes'', each subject to quantum fluctuations of its shape. The
\textit{shape quantum mechanics} of $p$-branes was discussed in detail in
previous papers \cite{PhD}.\\
Keeping in mind the scenario outlined above, in this 
paper we wish to introduce a new kind of quantum fluctuation.
Specifically, \textit{we will show that, at the Planck scale, not only shapes
and topology fluctuate, but the very
dimensionality ``$p$'' of a $p$-brane becomes a quantum variable}. \\
This result follows from an explicit calculation of the quantum
propagator of a bosonic $p$-brane in a new type of approximation 
\cite{queminpro}.
Let us recall that the world--history of an extended object is the
\textit{mapping} of a $p$-cell, i.e., a topological simplex of order ``$p$'',
to a physical $p$-brane, and is usually encoded into an action of
Dirac--Nambu--Goto type, or extensions of it\cite{firm}. From the action, in
principle, one may deduce the Green function by evaluating the path integral
over an  appropriate set of $p$-brane histories\cite{PhD}. In practice,
the quantum propagator of a bosonic $p$-brane cannot be computed
exactly in closed form. Thus, in a recent paper \cite{queminpro}
we borrowed the \textit{ minisuperspace quantization scheme} from
Quantum Cosmology \cite{mini} and the \textit{  quenching approximation}
from QCD \cite{quench} in order to derive a new form of the bosonic
$p$-brane propagator.\\
The  "minisuperspace approximation" is currently used to quantize 
restricted families of 
cosmological models. Invariance under general coordinate transformations is 
broken down to the restricted symmetry under coordinates changes 
that preserve the form of the cosmic line element. We have extended this 
approach from four dimensional
spacetime to the $p+1$ dimensional world manifold of the $p$-brane. Selecting
a restricted form of the world manifold metric ( not to be confused with
the target spacetime metric ) allows one to separate the center of mass 
dynamics from the $p$-brane deformations. \\
'' Quenching '', on the other hand, is a non-perturbative approximation 
scheme currently 
implemented
in gauge theories. The system is quantized in a ( large ) ``box'' where only
long wavelength fluctuations are quantized. High frequency modes are ``frozen''.
For a Yang-Mills type gauge theory, dealing with
long wavelengths means an infrared, non-perturbative, regime.
 In this approximation $SU(N)$ gauge theory turns into
a sort of matrix quantum mechanics amenable to large-$N$ expansion.
The presence of extended objects in the large-$N$ spectrum of matrix Yang-Mills
models is an old problem with some new results \cite{nlarge}. In
\cite{queminpro} we reversed the above line of reasoning and quenched
the high frequency modes of the brane history in target spacetime.\\
Combining those two techniques, i.e., minisuperspace in the world manifold
and quenching in the target spacetime, 
leads to an expression for the propagator that describes the \textit{boundary 
dynamics} of the brane \cite{PhD} rather than the spectrum of the bulk small
oscillations. This new approximation includes both the center of mass
quantum motion in target spacetime as well as the \textit{collective
deformations} of the brane as described by transitions between different
(hyper-)volume quantum states. The latter quantum jumps account for
the volume variations induced by local deformations of the brane shape, and
generalize the ``areal quantum dynamics'', originally formulated by Eguchi
in the case of strings \cite{eguchi}, to the case of higher dimensional
objects. \\
The paper is organized as follows.\\
In Section 2 we review the essential factorization of the exact propagator
in the minisuperspace, quenched approximation. By analogy with the 
familiar case of the point-like particle, we extract the " dispersion
relation " for a $p$-brane and the line element of the underlying 
spacetime geometry.\\
Section 3 deals with the second result that we wish to communicate in this 
paper, namely, the extension of  Heisenberg's principle in ordinary
Quantum Mechanics to the Planck Uncertainty Principle which is anticipated
in String Theory.\\
Section 4 concludes the paper with a brief summary of our discussions.

\section{Minisuperspace, Quenched Propagator}

The propagator calculated within the combined approximation described in the
introduction  takes  the following form in the Feynman parametrization scheme:

\begin{equation}
\fl G_p \left(\, x - x _{0}\ , \sigma \,\right)={i\over 2M_0}\,
\int_0^\infty ds\, \exp\left\{\, i\, s \,  {M_0\over 2}\, (p+1)\,\right\}
K_{cm}\left(\, x - x _{0}\ ; s\,\right)\, K_p\left(\,  \sigma \ ; s
\,\right)\ .
\label{prop}
\end{equation}

Here, $K_{cm}\left(\, x - x _{0}\ ; s\,\right)$ represents the propagation
amplitude of the brane center of mass in a $D$-dimensional target spacetime
$\mathcal{M}$, from the initial position $ x_0$ to the final one, i.e., 
$x$:

\begin{equation}
K_{cm}\left(\, x - x _0\ ; s\,\right)= \left(\,{\pi M_0\over is
}\,\right)^{D/2} \exp \left[\, i\, {M_0\over 2s}\, \left(\, x-x_0\,\right)^2\,
\right]\ .
\end{equation}

This part of the complete propagator encodes the free motion of the object as
a whole.
When observed with  poor resolution the brane looks like an ``effective 
point particle'' of mass $M_0$ concentrated at the center of mass. $M_0$ 
is defined by:

\begin{equation}
M_0\equiv m_{p+1}\,V_p 
\end{equation}

where $m_{p+1}$ is the brane \textit{tension} and $V_p$ is the proper volume
of the $p$-dimensional boundary of the brane world manifold. The quantum
deformations of the object are \textit{globally} described by the ``volume 
propagator'' $K_p\left(\,  \sigma \ ; s \,\right)$; it  represents 
the transition amplitude between the vacuum, i.e., a zero volume state, 
and a
$1$-brane state of finite proper volume $V_p$\footnote{Within the quenching 
approximation, $p$--branes with different shapes, but equal volume, are 
considered to be 
physically equivalent.}:

\begin{equation}
\fl 
K_p\left(\,  \sigma \ ; s\,\right)=\left(\,{ M_0\over i\pi V_p^2 s
}\,\right)^{{1\over 2}{D\choose p+1} }
\exp \left[\, {i\over (p+1)!}\, {M_0\over 2s V_p^2}\,
\sigma^{\mu_1\dots\mu_{p+1}} \, \sigma_{\mu_1\dots\mu_{p+1}}   \,\right]\ .
\label{kp}
\end{equation}

Let us denote by $y^\mu\equiv y^\mu\left(\, u^1\ , u^2\ , ...\  ,u^p\,\right)$ 
 the embedding functions of the world manifold boundary; then,  the volume 
multivector

\begin{equation}
\sigma^{\mu_1\dots\mu_{p+1}}\equiv \int y^{\mu_1}\, dy^{\mu_2}\wedge\dots
\wedge dy^{\mu_{p+1}}\ ,\qquad p\ge 1
\end{equation}

describes the target spacetime volume  enclosed by the brane.\\
Notice that the vacuum-to-vacuum propagator, i.e. the transition amplitude 
between two ``poinlike branes'', is simply a Dirac delta-function

\begin{equation}
\lim_{V_p\to 0}\, K_p\left(\,  \sigma 
 \ ; s\,\right)=\delta\left(\, \sigma\,\right)\ .
 \end{equation}

It may be helpful, before proceeding further with our discussion, to get a
``feeling'' of the quantum propagator by checking the consistency of our
formulation against some well known cases of physical interest. The precise
details of the derivation are in Ref.\cite{queminpro}.\\
The expression (\ref{prop}) of  the minisuperspace, quenched, propagator
holds for any $D$ and $p\ge 0$. Two special cases will illustrate how the
``machine'' works.\\
As we mentioned in the introduction, the model manifold of a 
pointlike particle is a $0$-cell  of vanishing proper volume $V_0=0$.
The form of (\ref{kp}) is simply

\begin{equation}
K_{p=0}\left(\,  \sigma \ ; s\,\right)=\delta\left(\,0\,\right)=1
\end{equation}

and the full propagator reduces to the familiar one:
\begin{equation}
\fl
G_{p=0 }\left(\, x - x _{0}\ , 0 \,\right)=
{i\over 2M_0}\,
\int_0^\infty ds\, \exp\left(\, i\, s \,  {M_0\over 2}\,\right)
\left(\,{\pi M_0\over i\, s}\,\right)^{D/2} \exp \left[\, i\, {M_0\over 2s}\,
\left(\, x-x_0\, \right)^2\,\right] \ .
\label{p0}
\end{equation}

In our description, a  string represents the target spacetime
image of a collection $\gamma$ of $1$-cells. Thus, the volume multivector 
collapses into the \textit{loop--coordinates} \cite{eguchi}

\begin{equation}
 \sigma^{\mu\nu}\equiv \oint_\gamma y^\mu\, dy^{\nu}
\end{equation}

while $V_1\equiv l_\gamma $ is  the string \textit{proper length}.
The resulting propagator takes the following form

\begin{eqnarray}
\fl G_p \left(\, x - x _{0}\ , \sigma \,\right)=    {i\over 2M_0}\,
\int_0^\infty ds \, \exp\left(\, i\, s \,  M_0\, \,\right)
K_{cm}\left(\, x - x _{0}\ ; s\,\right)\,
\left(\,{ M_0\over i\pi ^2 l_\gamma^2 s}\,\right)^{{1\over 2}{D\choose 2} }
\times\nonumber\\
 \lo \exp \left[\, {i\over 2}\, {M_0\over s\, l_\gamma^2}\, 
\sigma^{\mu\nu}\, \sigma_{\mu\nu}\,\right]\,
\label{p1}
\end{eqnarray}
with $M_0=l_\gamma\,m_2 \equiv l_\gamma/2\pi\alpha^\prime$.\\
Returning now to the general case, it is convenient to rewrite the
expression (\ref{prop}) in  ``~momentum space~''. By Fourier transforming 
(\ref{prop}) with the center of mass momentum $k_\mu$ and volume momentum 
$K_{\mu_1\dots\mu_{p+1}}$, we obtain :

\begin{eqnarray}
\fl G_p \left(\, x - x _{0}\ , \sigma \,\right)=
\int \frac{d ^{D} k}{ (2 \pi) ^{D}}\exp\left[\, i\, k_\mu\,(\, x - x _0\,)^\mu
\,\right]\times \nonumber\\
\lo  \int [ d K ] \exp\left[\,
{i\over (p+1)!}\, K_{\mu_1\dots\mu_{p+1}} \, 
\sigma^{\mu_1\dots\mu_{p+1}}\,\right]{1\over k ^{2}+(p+1)\, M_0^2+{V ^{\, 2}
_{p}\over (p+1)!}\, K_{\mu_1\dots\mu_{p+1}}^2 }\ .\nonumber\\
\label{Green}
\end{eqnarray}

where $\int [ d K ] \left(\, \dots\, \right) $ is a generalized gaussian 
integral.\\
Inspection of the Feynman--Green function (\ref{Green}) reveals a strong
similarity with the propagator of a ``pointlike'' probe, 
provided one interprets the pair of conjugate
volume variables $\left(\, \sigma\ , K \,\right)$ as
new ``coordinates'' in an extended phase space \cite{ogiel}. \\
One may regard the expression (\ref{Green}) as the propagator of a 
generic $p$-brane combined with its barycentric
coordinate, or as a pointlike probe moving in a spacetime with 
metric \cite{pez1} :

\begin{equation}
    d \Sigma ^{2}
    =
    dx ^{\mu}\, d x _{\mu}
    +
    \frac{1}{(p+1)! V _{p} ^{\, 2}}\,
    d\sigma^{\mu _{1} \dots \mu _{p+1}}\,
    d\sigma_{\mu _{1} \dots \mu _{p+1}}\ .
\label{eq:linelm}
\end{equation}

The key result of this paper is the \textit{equivalence} between the
quantum propagator  (\ref{Green}) and the ``classical'' motion of a
pointlike probe in the geometry defined by the line element (\ref{eq:linelm}).
In the string case, i.e., $p=1$,
the size of the areal corrections is determined by the proper length $l_\gamma$.
In the general case the constitutive elements of the geometry (\ref{eq:linelm})
are smeared over a finite volume which is the target spacetime image of 
the proper volume $V _{p}$.\\ 
Whenever $V ^{\, 2} _{p}\, K^2<< k^2$, ``the volume modes'' are not 
excited and the propagator reduces to a vanishing volume form 

\begin{eqnarray}
\fl G_p \left(\, x - x _{0}\ , \sigma \,\right)\approx
\int \frac{d ^{D} k}{ (2 \pi) ^{D}}\exp\left[\, i\, k_\mu\,(\, x - x _0\,)^\mu
\,\right]\times \nonumber\\
\lo  \int [ d K ] \exp\left[\,
{i\over (p+1)!}\, K_{\mu_1\dots\mu_{p+1}} \, 
\sigma^{\mu_1\dots\mu_{p+1}}\,\right]\,
 {1\over k ^{2}+(p+1)\, M_0^2}
\nonumber\\
\lo =\delta\left(\, \sigma \,\right)
\int \frac{d ^{D} k}{ (2 \pi) ^{D}}\exp\left[\, i\, k_\mu\,(\, x - x _0\,)^\mu
\,\right]{1\over k ^{2}+(p+1)\, M_0^2}
\label{G0}
\end{eqnarray}

Accordingly, the probe ''sees'' only the Minkowski sub-space of
(\ref{eq:linelm}). However, as soon as energy is transferred into the
volume modes, and $V ^{\, 2} _{p}\, K^2$ becomes comparable with 
$k^2$, new dimensional channels open up, because of the  extended nature 
of the object.
 Alternatively, volume quantum transitions can be interpreted as an effect 
due to the new kind of geometry tested by the probe.
The metric element (\ref{eq:linelm})  corresponds to the $\left(\, 0\ , p 
\,\right)$ sub-space of the \textit{Clifford line ``element''}:

\begin{eqnarray}
       \fl    d \Sigma ^{\, 2}
    =
    d x ^{\mu}\,
    d x _{\mu}
    +
    \frac{1}{2\, l  ^{2}}\,
    d  \sigma ^{\mu _{1} \mu _{2}}\,
    d  \sigma _{\mu _{1} \mu _{2}}
    +
    \frac{1}{3!\, l^{\, 4}}\,
    d  \sigma ^{\mu _{1} \mu _{2} \mu _{3}}\,
    d  \sigma _{\mu _{1} \mu _{2} \mu _{3}}
    +
    \dots
    +
    \nonumber \\
   \lo    
    +
    \frac{1}{(p+1)!\, l^{\, 2p}}\,
    d  \sigma ^{\mu _{1} \dots \mu _{p+1}}\,
    d  \sigma _{\mu _{1} \dots \mu _{p+1}}
    +
    \dots
    +
    \frac{1}{D!\,  l^{\, 2(D-1)}}\,
    d  \sigma ^{\mu_1 \dots \mu_D}\,
    d  \sigma _{\mu_1 \dots \mu_D}\ .
    \label{poly}
\end{eqnarray}

The line element (\ref{poly}) takes into account the whole range $0\le 
p\le D-1$.
As a mathematical construct, the line element (\ref{poly})  has long been
known, at least to some  practitioners of Clifford algebras. For instance,
Pezzaglia has introduced it to discuss the long standing problem of a
classical spinning particle \cite{pez1}\cite{pez2}, while Castro and Pavsic
studied both higher order derivative gravity and ordering ambiguity in this
geometric background \cite{pavsic}. In this new context, Eq.
(\ref{poly}) may be interpreted as an extension of the usual Lorentzian
line element. \\
To our mind, however, the Clifford line element establishes a mathematical
and physical link to the ``unified theory of relativistic extended
objects'', which has been under consideration by 
 the authors for a number of years. Indeed, the whole cardinal
concept of \textit{relativity of motion} may be extended to the broader
context of \textit{relativity of dimensions.} By ``relative
dimensionalism'', we mean that the new Clifford metric opens the possibility of 
transformations between different $p$-branes, so that their
effective dimensionality, and the very geometry of spacetime, becomes
\textit{resolution dependent.} \\
The \textit{polydimensional propagator} corresponding to the general
line element (\ref{poly}) takes the form 

\begin{eqnarray}
\fl G_{poly.} \left(\, x - x _{0}\ , \sigma \,\right)=
\int \frac{d ^{D} k}{ (2 \pi) ^{D}}\exp\left[\, i\, k_\mu\,(\, x - x _0\,)^\mu
\,\right]\times \nonumber\\
\lo \Pi_{p=1}^{D-1}\left\{\, \int [ {\mathcal{D}} K ]_p \exp\left[\,
{i\over (p+1)!}\, K_{\mu_1\dots\mu_{p+1}} \, \left(\,
\sigma^{\mu_1\dots\mu_{p+1}}\,
\right)\,\right]\, \right\}\times\nonumber\\
 {1\over k ^{2}+ M^2_{poly.}+\sum_{p=1}^{D-1}\left[\, 
{V ^{\, 2} _{p}\over (p+1)!}\, K_{\mu_1\dots\mu_{p+1}}^2\,\right]}\ .
\label{gp}
\end{eqnarray}

The object represented by (\ref{gp}) is a \textit{ quantum superposition 
of diverse-$p$ branes.}  We  call it \textit{ Planckion,} because
its length scale will be shown in the next section to be comparable with
the Planck Length. The Planckion is the target spacetime 
mapping of the whole  simplicial complex of elementary $p$-cells.\\
In this new framework, there is no essential difference among 
objects with different dimensions, there is only one fundamental object,
the Planckion.
In euclidean three dimensional space, one can rotate the frame of 
reference and set to zero one or two components of an ordinary vector. 
Similarly, a  ``rotation'' in the 
Clifford geometry (\ref{poly}) allows to project the  Planckion on
a sub-space of definite dimension. By selecting a definite-$p$ classical 
action one picks up the $\left(\, 0\ ,p\,\right)$ component of the 
Planckion. In order to reproduce the full
propagator (\ref{gp}) one should start from a classical action of the form
\cite{queminpro}

\begin{equation}
S_{Pl}=S_{cm}-\frac{1}{2}\sum_{p=1}^{D-1}\left\{\, \int d^{p+1}\sigma
\sqrt{\gamma_p}\left[\, \gamma_p^{mn}\,\partial_m\, Y^\mu_p\,\partial_n\,
 Y_{\mu\,p} -p\, m_{p+1}\,\right] \,\right\}
\label{spoly}
\end{equation}

where $S_{cm}$ is the action of the center of mass of a polydimensional
brane. The quantization of the classical model (\ref{spoly}), leading 
to the propagator (\ref{gp}), will be discussed in a separate paper.
In the next section we shall justify the choice of the term  Planckion   
by showing how the proper volume
$V_p$, or the fundamental scale $l \quad \left(\, =\left[\, V_p\,\right]^{1/p} 
\,\right)$, is related  to the Planck length 
through a generalized Uncertainty Principle.

\section{Generalized Uncertainty Principle}

In order to substantiate the  connection between the Clifford line element
and the short distance structure of spacetime, it is worth recalling
that the generalized Lorentzian metric (\ref{poly})  calls for
the introduction of a fundamental length, $l$, or  energy scale $ l^{-1}$,
in the fabric of spacetime.  The distinct novelty, and advantage, of 
the Clifford line element is that  scalars, vectors,
bivectors, or $p$-vectors can be summed together in the same metric formula, 
or, in brane language,  
a definite-$p$ object unfolds its basic polydimensional nature over distances
comparable with $l$.  \\
In order to clarify the physical meaning of the fundamental scale
$l$, let us consider again the $\left(\, 0\ , p\,\right)$ component of the 
Planckion propagator.
The new \textit{   tension--shell} condition, or " dispersion relation ", 
defined by the vanishing of the denominator in Eq.(\ref{Green}) is:

\begin{equation}
k^2 + {V_p^2\over (p+1)!}
K^2_{\mu_1\dots\mu_{p+1}} +(p+1)\, M_0^2=0\ .
\label{disp}
\end{equation}

Real branes satisfy the constraint (\ref{disp}) which correlates center of
mass and volume momentum squared, which are independent quantities for 
\textit{virtual} branes. Let us further parametrize the square of the volume
momentum in terms of the modulus of the center of mass four momentum
\footnote{
\begin{equation}
\left[\, k_\mu\,\right]=(\,{\mathrm{length}}\,)^{-1}\ ,\qquad
\left[\, k_{\mu_1\dots\mu_{p+1}}\,\right]=(\,{\mathrm{length}}\,)^{-(p+1)}
\nonumber
\end{equation}}:

\begin{equation}
 {1\over (p+1)!}\, K^2_{\mu_1\dots\mu_{p+1}}\equiv \beta\,
k^2 ( k^2 )^p\ ,
\end{equation}

We expect the volume modes to be excited only when the center of mass
wavelength becomes comparable with the length scale of the volume $V_p$.
Accordingly, the proportionality constant $\beta$ is expected to be an $0(1)$
number. Thus, the tension shell
condition (\ref{disp}) can be written as a $(p+1)$-degree polynomial in
$k^2$. It is convenient, at this point, to restore the explicit presence of
Planck's constant in the above equation. Thus, the ``tension shell condition''   
can be cast in the more familiar form of mass--shell condition for a 
pointlike particle by incorporating the extra $k^2$-dependence within an 
effective, or \textit{running Planck constant} \cite{castro}
\begin{equation}
\fl
\hbar^{{\mathrm{eff.}}}\equiv \hbar\, \left[\, 1 + V_p^2\, \beta\, 
(\, k^2\,)^p
\,
\right]^{1/2} \qquad \longrightarrow\qquad \left(\, \hbar^{{\mathrm{eff.}}} k
\,\right)^2
+(p+1)\, M_0^2=0\ .
\end{equation}

From here follows the generalized form of the Uncertainty Principle :

\begin{equation}
\Delta x^\mu\, \Delta k_\mu\ge \langle\, \hbar^{{\mathrm{eff.}}}\,\rangle
\ ,
\label{genh}
\end{equation}

where $\langle\,\dots \,\rangle $ denotes vacuum average.\\
The physical content of (\ref{genh}) emerges in the non-relativistic limit,
in which case we drop out $0$-components, 
and rotate the reference frame in such a way that the $x$-axis
coincides with the direction of the center of mass position vector $\vec x$:

\begin{equation}
\Delta x\, \Delta  k_x\ge \langle\, \hbar^{{\mathrm{eff.}}}\,
\rangle \approx {\hbar\over 2}\, \left(\, 1 + {1\over 2}\,
V_p^2\, \beta\, \left(\,\langle k_x^2\,\rangle\right)^p + \dots
\right)\ .
\label{nrelx}
\end{equation}

Finally, by taking into account  the translational invariance of the
vacuum, that is, $\langle \, k_x \,\rangle =0$, we can relate momentum
uncertainty and momentum fluctuation as follows,
$\Delta  k_x\equiv \sqrt{\,\langle \, k_x^2\,\rangle - \langle
k_x\,\rangle^2}=\sqrt{\,\langle \, k_x^2\,\rangle } $.

Thus, the position uncertainty reads

\begin{equation}
\Delta  x \ge  {\hbar\over \Delta k_x}\, \left[\, 1 + {1\over 2}\,
V_p^2\, \beta\, \left(\,\Delta k_x \,\right)^{2p} + \dots  \right]\ .
\label{nrel2}
\end{equation}

\textit{  This formula holds for any $p$,} and therefore it is applicable to
strings as well. Thus, comparing with the ``string
uncertainty principle'' \cite{mende}

\begin{equation}
\Delta  x \ge  {\hbar\over \Delta k_x}\, \left[\, 1 + \,
\, {\mathrm{const.}}\times\, G_N\left(\,\Delta k_x \,\right)^2\, \right]
\ ,
\label{string}
\end{equation}

we find that the proper volume of the collection of $p$-cells is proportional 
to the Planck volume

\begin{equation}
V_p\propto\left(\, l_{Pl}\, \right)^p
=\left(\,\hbar \, G_N\, \right)^{p/2}\ .
\end{equation}

The standard procedure, at this point, is to minimize the position uncertainty

\begin{equation}
{d ( \Delta  x) \over d ( \Delta k_x)}=0\ \Longrightarrow
\left(\, \Delta k_x\,\right)^2_{min}=\left[\, {2\over (2p-1)\,
\,\beta\,\left(\,\hbar \, G_N\, \right)^p}   \, \right]^{1/p}\ , \qquad p\ge 1
\label{kmin}
\end{equation}

so that, by inserting (\ref{kmin}) in (\ref{nrel2}), one gets

\begin{equation}
\left(\,\Delta  x\,\right)_{min}\propto \hbar\,l_{Pl}
\ .
\end{equation}
%By switching once again to  natural units, $\hbar=1$, we see that the
%minimum uncertainty, or \textit{quantum of resolution} follows from 
%the   length scale $V_p^{1/p}$, i.e.
%$\left(\,\Delta  x\,\right)_{min}\propto l_{Pl}$ .
According to this fundamental condition,  strings, or any
$p$-brane, or test body for that matter, cannot probe distances shorter
than $l_{Pl}$. Thus, the length scale $l$, originally introduced in the
line element (\ref{poly}) for purely dimensional reasons, can now be related
to a \textit{minimum universal length}\footnote{The necessity and
role of a minimum length in quantum gravity is reviewed in Ref.\cite{garay}.}.

\section{Discussion and Conclusions}

The main results of this paper can be summarized as follows. 
We started from a new type of Minisuperspace  propagator encoding both the 
motion of a bosonic $p$-brane as a whole, and its volume evolution. 
The spacelike volume of the brane, expressed in terms of the
multivector $\sigma^{\mu_1\dots\mu_{p+1}}$, is the mapping of the  
world manifold proper volume $V_p$. 
The ensuing  result can be interpreted  as
a standard propagator in a non-standard geometry in which
both vector $x^\mu$ and $\sigma^{\mu_1\dots\mu_{p+1}}$ play the role
of ``coordinates''. This non-standard geometry
is part of a Clifford line element where points, lines, volumes, can be summed
together thanks to the presence of a suitable constant with length dimension.
In our approach, this length scale is fixed by the form of the propagator,
i.e., $l=(\, V_p\,)^{1/p}$.  For $p=1$, $V_1$ is the string proper 
length and results to be proportional to the Planck Length $l_{Pl}$. 
The target spacetime counterpart of this fundamental length scale
is a minimum distance entering a generalized Uncertainty Principle 
derived from our  propagator. Thus, the Clifford line element 
naturally encodes the notion of minimum distance as a consequence of quantum
effects. On the other hand, the Principle of Dimensional Democracy implemented
in the Clifford geometry, provides a unification principle underlying
$p$-brane dynamics at the Planck scale.

\ack{ We are very much in debt to William Pezzaglia Jr. and Theodore G. Erler
IV  for many enlightening and stimulating conversations about the meaning and
possible applications of Clifford's Calculus.\\
Two of us, S.A. and E.S., would like to thank Prof.M. Pavsic and Prof.
C.P.Castro for some useful discussions during their visit at the
Department of Theoretical Physics of the University of Trieste.}

\end{document}